\documentclass[aps]{revtex4}
\usepackage[dvips]{epsfig}
\usepackage{amsmath,amssymb}

\parskip=\medskipamount

\newcommand{\eq}[1]{(\ref{#1})}

\newcommand{\be}{\begin{equation}}
\newcommand{\ee}{\end{equation}}

\newcommand{\barr}{\begin{array}}
\newcommand{\earr}{\end{array}}

\newcommand{\beqn}{\begin{eqnarray}}
\newcommand{\eeqn}{\end{eqnarray}}

\newcommand{\bs}{\begin{subequations}}
\newcommand{\es}{\end{subequations}}

\newcommand{\bw}{\begin{widetext}}
\newcommand{\ew}{\end{widetext}}

\newcommand{\mbf}{\mathbf}
\newcommand{\mcl}{\mathcal}

\newcommand\disp{\displaystyle}

\begin{document}

\title{Whether the mean--field two--length scale theory of hydrophobic effect can be
microscopically approved?}
\author{G. Sitnikov$^{1,2}$, S. Nechaev$^{3,4}$}
\affiliation{$^1$Moscow Institute of Physics and Technology, Institutskaya str.9,
141700, Dolgoprudnyi, Russia \\ $^2$Algodign LLC, Bolshaya Sadovaya str., 8(1),
123379, Moscow, Russia \\ $^{3}$LPTMS, Universit\'e Paris Sud, 91405 Orsay Cedex,
France \\ $^{4}$L.D.Landau Institute for Theoretical Physics, 117334 Moscow, Russia}

\date{October 3, 2005}

\begin{abstract}
We discuss the simple microscopic derivation of a hydrophobic effect. Our approach
is based on the standard functional representation of the partition function of
interacting classical particles and subsequent passage to collective variables
(local densities of the solvent). We get an expression for the solvation free energy
of solute molecule of any arbitrary shape and derive the nonlinear equation for the
mean solvent density surrounding the solvated object. We pay a special attention to
some inconsistencies between the microscopic consideration and the two--length scale
mean--field theory of hydrophobic effect.
\end{abstract}

\maketitle

\section{Introduction}
\label{sect:1}

The key problem in thermodynamics of hydrophobic interactions consists in accounting
of the effect of fluctuating media on interactions between solvated molecules. In
brief, the hydrophobic effect occurs due to expelling the solvent from the volume
occupied by solute molecule. Hence, the effect of solvation can be accounted by
forcing the total solvent density to be zero inside the solute molecule.

Among various theoretical attempts in developing a constructive theory of
solvent--solute interactions, special attention deserve the works where the
hydrophobic effect is treated in the mean--field approximation with explicit
separation of two characteristic length scales
\cite{chand_gauss,chand_LCW,chand_lattice}. In this approach the fluctuating solvent
density is decomposed in two components, the slowly varying field describing the
mean solvent (water) density, and the short--ranged density fluctuations (usually
supposed to be Gaussian) describing correlations in the solvent on scales of order
of size of solvent molecules. There is a general belief that such an approach is
optimal from different points of view: on one hand it is physically clear, being
"semi-microscopic", and on other hand, it can be used as a constructive
computational tool of account of water, much faster than corresponding explicit
approaches, but without essential loss of precision. In particular, minimization of
the free energy in the frameworks of the two--length scale approach allows to
determine the structure of hydrophobic layer and the solvation free energy of solute
molecules of any geometry, if the density--density correlation function of the pure
solvent is known.

To be precise, the two--length scale approach deals with the following Hamiltonian
\be
\mcl{H}_0[\omega,n]=\frac{1}{2}\int \omega({\bf r})
\chi^{-1}({\bf r},{\bf r}') \omega({\bf r}') \, d{\bf r} d{\bf r}' + \int
\left\{\frac{a}{2} \left(\nabla n({\bf r})\right)^2+ W(n({\bf r}))\right\}d{\bf r} +
c\int n({\bf r}) \omega({\bf r})\, d{\bf r} \label{1}
\ee
where $n({\bf r})$ is the smoothly changing (average) solvent density; $\omega({\bf
r})$ is the field corresponding to the short--ranged density fluctuations;
$\chi(|{\bf r}-{\bf r}'|)$ is the solvent correlation function in the bulk; $a$ is
the phenomenological parameter which requires the microscopic determination---see
the discussions in \cite{Nechaev&Sitnikov&Taran}; the last term describes the
interaction between short-- and long--scale terms with the coupling constant $c$;
and the self--consistent potential $W(n({\bf r}))$ is chosen in the common form of
the standard Ginzburg--Landau (GL) expansion for the order parameter $n({\bf r})$ as
the fourth--order polynomial allowing the liquid--vapor phase transition:
\be
W(n)= \frac{b}{2} (n-n_1)^2\,(n-n_2)^2 \qquad (0\le n_1\le n_2\le 1) \label{eq:GLP}
\ee
where $n_1$ and $n_2$ are the values of the order parameter $n$ in the vapor and
liquid phases correspondingly (below we set, if not specified, $n_1=0$) and $b$ is
the coupling constant which in combination with the parameter $a$ defines the
surface tension.

Following the general scheme of the works \cite{chand_lattice}, we suppose that the
solvent cannot penetrate into the volume occupied by the solute molecule. Hence the
total solvent density $\rho({\bf r})=n({\bf r})+\omega({\bf r})$ is nullified inside
the solute: $\rho({\bf r})=0$ for all ${\bf r}\in v_{\rm in}$, where $v_{\rm in}$ is
the volume occupied by the solute molecule. The fact that we force the total solvent
density $\rho({\bf r})=n({\bf r})+\omega({\bf r})$ to be zero inside the solute,
results in an effective interactions between $n({\bf r})$ and $\omega({\bf r})$. In
general, one can permit also the direct coupling between $n({\bf r})$ and
$\omega({\bf r})$ everywhere in the solute. Minimizing the corresponding free energy
functional, we arrive at the set of equations for the profile of the mean solvent
density $n({\bf r})$ in presence of the solute molecule:
\bs \beqn -a \Delta n({\bf
r}) + \frac{\delta W(n({\bf r}))}{\delta n({\bf r})}+ \int_{v_{\rm in}}d{\bf r}'
\chi^{-1}_{\rm in}({\bf r},{\bf r}') n({\bf
r}') = 0 & \qquad \mbox{for ${\bf r}\in v_{\rm in}$} \label{2a} \\
\disp -a \Delta n({\bf r}) + \frac{\delta W(n({\bf r}))}{\delta n({\bf r})} + U({\bf
r}) = 0 & \qquad \mbox{for ${\bf r}\in v_{\rm out}$} \label{2b}
\eeqn \es
where it is supposed for simplicity that $c=0$ (see \eq{1}). We are able to compute
also the corresponding solvation free energy---see \cite{Nechaev&Sitnikov&Taran} for
details.

The theory of hydrophobic effect based on the two--scale Hamiltonian \eq{1} with
properly adjusted parameters $a$ and $b$ describes qualitatively and even
quantitatively many physical effects. For example, it reproduces the oscillatory
behavior in the free energy of interactions of two solute molecules as the function
of their mutual distance, and gives with good precision the experimentally measured
values of solvation free energy of alkane molecules \cite{Nechaev&Sitnikov&Taran}.

However the Hamiltonian \eq{1} is written {\it ad hoc} on the basis of physical
suppositions on the possibility of separation of interactions in "short" and "long"
scales. If the two--length scale theory pretends to describe all the peculiarities
of the hydrophobic effect, it should be confirmed by more solid arguments. For
example, one could ask a question whether such a theory can be approved
microscopically. Just this question is addressed in our paper. The results of our
attempts to derive the Hamiltonian \eq{1} from microscopic consideration are
presented in Section \ref{sect:3}.

\section{From microscopic Hamiltonian to density functional}
\label{sect:2}

To make a contents of the paper as self--contained as possible, we describe in this
section basic steps of passing from microscopic description of classical system with
binary interactions to the pre-averaged description in terms of collective
variables. The hydrophobic effect will be considered in Section \ref{sect:3}.

\subsection{Collective variables}
\label{sect:2a}

We begin with a partition function of a grand canonical ensemble for the system of
$n$ identical classical particles
\be
Z=\sum_{n=0}^{\infty} \frac{\lambda^n}{n!}\int \left(\prod_{i=1}^{n}
d\mbf{r}_i\right) e^{-\beta H_{n}} \label{PF_particle}
\ee
The Hamiltonian $H_n$ of binary interacting particles has a standard form:
\be
H_{n}= \sum_{i=1}^{n} U(\mbf{r}_i)+\sum_{i<j}^{n}V(\mbf{r}_i-\mbf{r}_j)
\label{mic_ham}
\ee
where $U(\mbf{r})$ is an external field and $V(\mbf{r}-\mbf{r}')$ is a pairwise
potential. The activity $\lambda$ is related to the chemical potential $\mu$ by the
relation
$$
\lambda=\xi\, e^{\beta\mu},\quad \xi=\frac{2\pi m kT}{h^2}
$$
and $\beta=1/(kT)$ is the inverse temperature.

Let us define the density
$$
\rho(\mbf{r})=\sum_{i=1}^{n} \delta(\mbf{r}-\mbf{r}_i)
$$
Now we can pass from coordinates of individual particles, ${\bf r}_i$ ($i=1,...,
n$), to the pre-averaged local collective variables, $\rho(\mbf{r})$, via the
standard technique of constraints in the functional integral \cite{Negele&Orland}.
Using the relation
$$
\int \mcl{D}\{\rho\}\, \delta\Big{[}\rho(\mbf{r})-\sum_{i=1}^{n}
\delta(\mbf{r}-\mbf{r}_i)\Big{]}=1
$$
and the functional Fourier transform
\be
\delta\Big{[}\rho(\mbf{r})-\sum_{i=1}^n \delta(\mbf{r}-\mbf{r}_i)\Big{]}=\int
\mcl{D}\{\phi\}\, \exp\left\{i\int \phi(\mbf{r}) \left(\rho(\mbf{r})-\sum_{i=1}^n
\delta(\mbf{r}-\mbf{r}_i)\right) d\mbf{r}\right\}
\ee
we get
\begin{align}
\disp Z=\sum_{n=0}^{\infty}\int \mcl{D}\{\phi\}\, \frac{\lambda^{n}}{n!} & \disp
\int \left(\prod_{i=1}^{n}d\mbf{r}_i\right)\, \exp\left\{-i\beta \sum_{i=1}^{n}
\phi(\mbf{r}_i)\right\} \nonumber \\ & \disp \times\int \mcl{D}\{\rho({\bf r})\}\,
\exp\left\{i\beta \int d\mbf{r}\,\rho(\mbf{r})\phi(\mbf{r})-\beta \int d\mbf{r}\,
\rho(\mbf{r}) U(\mbf{r})-\frac{\beta}{2}\int d\mbf{r}\, d\mbf{r}' \,
\rho(\mbf{r})V(\mbf{r}-\mbf{r}') \rho(\mbf{r}')\right\}
\end{align}
Carrying out the summation over $n$ we obtain
\be
Z=\int\int \mcl{D}\{\rho\}\mcl{D}\{\phi\}\, e^{-S(\rho,\phi)}
\ee
where
\be
S\{\rho,\phi\}=\frac{\beta}{2} \int \rho(\mbf{r})V(\mbf{r}-\mbf{r}') \rho(\mbf{r}')
d\mbf{r}d\mbf{r}'-\beta \int \rho(\mbf{r}) \Big(\phi(\mbf{r})-U(\mbf{r})\Big)
d\mbf{r} -\lambda \int e^{-\beta \phi(\mbf{r})}d\mbf{r}
\ee
and $\phi$ rotated to the imaginary axis.
After the functional integration over the field $\rho({\bf r})$, we have
\be
Z=\mcl{N}\int \mcl{D}\{\phi\}\, e^{-\tilde{S}(\phi)}
\ee
where
\be
\tilde{S}\{\phi\}=-\frac{\beta}{2}\int\Big(\phi(\mbf{r})-U(\mbf{r})\Big) \,V^{-1}
(\mbf{r}-\mbf{r}') \Big(\phi(\mbf{r}')-U(\mbf{r}')\Big) d\mbf{r}d\mbf{r}'-\lambda
\int\,e^{-\beta \phi(\mbf{r})}d\mbf{r}
\ee
The equilibrium value of the field $\phi$ can be directly obtained by minimizing the
effective action, $\tilde{S}$, of the system
\be \left.\frac{\delta \tilde{S}\{\phi\}}{\delta \phi}\right|_{\,\phi=\bar{\phi}}=
\beta \lambda e^{-\beta \bar{\phi}(\mbf{r})}-\beta\int V^{-1}(\mbf{r}-\mbf{r}')
\left(\bar{\phi}(\mbf{r}')-U(\mbf{r}')\right) d\mbf{r}'=0 \label{avr_phi}
\ee
The mean density $\bar{\rho}(\mbf{r})$ is determined now by
\be
\bar{\rho}(\mbf{r})=-\frac{1}{\beta} \frac{\delta \ln Z}{\delta U(\mbf{r})}
\ee
that reads
\be
\bar{\rho}(\mbf{r})=\int V^{-1}(\mbf{r}-\mbf{r}')\left(
\bar{\phi}(\mbf{r}')-U(\mbf{r}')\right) d\mbf{r}' \label{avr_rho}
\ee
After a little algebra we arrive at the nonlinear self--consistent equation for the
average density $\bar{\rho}$
\be
\bar{\rho}(\mbf{r})=\lambda\, \exp\left\{-\beta\left(U(\mbf{r})+\int
V(\mbf{r}-\mbf{r}')\bar{\rho}(\mbf{r}') d\mbf{r}'\right)\right\} \label{sce_rho}
\ee
The parameter $\lambda$ and hence, $\mu$, is determined by the value of $\bar{\rho}$
in the bulk, $\rho_{\rm b}$. Rewritten in the terms of bulk density \eq{sce_rho}
reads
\be
\bar{\rho}(\mbf{r})=\rho_{\rm b} \exp\left\{-\beta\left( U(\mbf{r})+\int
V(\mbf{r}-\mbf{r}')\left(\bar{\rho}(\mbf{r}') -\rho_{\rm
b}\right)d\mbf{r}'\right)\right\}
\ee

\subsection{Free energy functional}
\label{sect:2b}

Equations \eq{avr_phi}, \eq{avr_rho} and \eq{sce_rho} are sufficient to obtain an
explicit expression of $\bar{\rho}$ and of thermodynamic functionals $\Omega$ and
$F$. In the saddle point approximation the grand thermodynamic potential $\Omega$ is
\be
\Omega=-\frac{1}{\beta} \ln Z=\frac{S\{\bar{\phi}\}}{\beta}
\ee
Using \eq{avr_rho} and \eq{sce_rho}, we obtain
\be
\Omega=-\frac{1}{2}\int \bar{\rho}(\mbf{r})V(\mbf{r}-\mbf{r}')\bar{\rho}(\mbf{r}')
d\mbf{r} d\mbf{r}'-\frac{1}{\beta}\int \bar{\rho}(\mbf{r}) d\mbf{r}
\label{Eq_omega_rho}
\ee
From \eq{sce_rho} we can easily get the chemical potential $\mu=\frac{1}{\beta}\ln
\lambda + {\rm const}$
\be
\mu=\frac{1}{\beta} \ln \bar{\rho}(\mbf{r})+U(\mbf{r})+\int V(\mbf{r}-\mbf{r}')
\bar{\rho}(\mbf{r}') d\mbf{r}' + {\rm const} \label{Eq_mu_rho}
\ee
(we can remove the constants in \eq{Eq_mu_rho}, because they determine only the
reference state and do not give any contribution to physically important
quantities). Recalling that
\be
F=\Omega+\int \mu(\rho) \rho(\mbf{r}) d\mbf{r}
\ee
we arrive at the following expression for the free energy functional, $F$
\be
F=\frac{1}{2}\int \bar{\rho}(\mbf{r}) V(\mbf{r}-\mbf{r}')
\bar{\rho}(\mbf{r}')d\mbf{r}d\mbf{r}'+\frac{1}{\beta}\int
\bar{\rho}(\mbf{r})(\ln\bar{\rho}(\mbf{r})-1) d\mbf{r} +\int U(\mbf{r})
\bar{\rho}(\mbf{r}) d\mbf{r}\label{free_energy}
\ee
The grand canonical potential $\Omega$ determined by \eq{Eq_omega_rho} depends on
density distribution $\bar{\rho}(\mbf{r})$ only, while from the "thermodynamic
viewpoint" it should be the function of $\mu$. Taking into account that according to
\eq{Eq_mu_rho} the chemical potential $\mu$ in \eq{Eq_omega_rho} is expressed
already as a function of $\bar{\rho}$, we can reconstruct the thermodynamically
consistent form of $\Omega$. Starting from the free energy functional
\eq{free_energy} and applying Legendre transform, we obtain the grand canonical
potential with explicit dependence on $\mu$:
\be
\Omega=\frac{1}{2}\int \bar{\rho}(\mbf{r}) V(\mbf{r}-\mbf{r}')
\bar{\rho}(\mbf{r}')d\mbf{r}d\mbf{r}'+\frac{1}{\beta}\int
\bar{\rho}(\mbf{r})(\ln\bar{\rho}(\mbf{r})-1) d\mbf{r} +\int U(\mbf{r})
\bar{\rho}(\mbf{r}) d\mbf{r}-\mu \int \bar{\rho}(\mbf{r}) d\mbf{r}  \label{Big_pot}
\ee

\subsection{Correlation and response functions}
\label{sect:2c}

There are several equivalent ways that allow to obtain the correlation function from
density functional \cite{Rowlinson_Widom}. We use the procedure based on variation
of the density of a liquid with respect to an external potential $U(\mbf{r})$.
According to \cite{Rowlinson_Widom} the one--point (direct) correlation function
$c^{(1)}({\bf r})$ can be defined via the following relation
\be
\rho(\mbf{r})=\lambda \,\exp\left\{-\beta U(\mbf{r})+c^{(1)}(\mbf{r})\right\}
\label{eq:c1}
\ee
Comparing \eq{eq:c1} and \eq{sce_rho}, we arrive at the following explicit
expression for $c^{(1)}({\bf r})$:
\be
c^{(1)}(\mbf{r})=-\beta \int V(\mbf{r}-\mbf{r}') \bar{\rho}(\mbf{r'}) d\mbf{r}'
\ee
Exploiting standard relation (see, for example, \cite{Rowlinson_Widom})
\be
c^{(2)}(\mbf{r},\mbf{r}')=\frac{\delta c^{(1)}(\mbf{r})}{\delta \rho(\mbf{r}')}
\ee
we get
\be
c^{(2)}(\mbf{r},\mbf{r}')=-\beta V(\mbf{r}-\mbf{r}')
\ee
here $c^{(2)}(\mbf{r})$ is the well-known Orstein--Zernike direct correlation
function. The inverse response function $\chi(\mbf{r},\mbf{r}')^{-1}$ reads
\be
\chi(\mbf{r},\mbf{r}')^{-1}=\frac{\delta(\mbf{r}-\mbf{r}')}{\rho(\mbf{r})}-c^{(2)}
(\mbf{r},\mbf{r}')= \frac{\delta(\mbf{r}-\mbf{r}')}{\rho(\mbf{r})}+\beta
V(\mbf{r}-\mbf{r}') \label{corr_fn}
\ee

The connection to experimentally accessible quantities is established by means of
the following equations
\be
\chi(|\mbf{r}-\mbf{r}'|)=\rho_{\rm b} \delta(\mbf{r}-\mbf{r}')+\rho_{\rm b}^2
h(|\mbf{r}-\mbf{r}'|);\qquad \int
\chi(|\mbf{r}-\mbf{r}''|)\chi^{-1}(\mbf{r}'',\mbf{r}') d\mbf{r}''=
\delta(\mbf{r}-\mbf{r}') \label{exp_corr_func}
\ee
where $h(\mbf{r})=g(\mbf{r})-1$ is the experimentally measurable correlation
function of the pure liquid.

Let us mention that the results \eq{corr_fn}--\eq{exp_corr_func} can be obtained by
exploiting an exact relation known as the first Yvon equation \cite{Rowlinson_Widom}
and \eq{exp_corr_func}, i.e.
$$
h(\mbf{r},\mbf{r}')=\frac{-kT}{\rho(\mbf{r})\rho(\mbf{r}')} \frac{\delta
\rho(\mbf{r})}{\delta U(\mbf{r}')}-\frac{\delta(\mbf{r}-\mbf{r}')}{\rho(\mbf{r}')}
$$

In case of more sophisticated form of entropy term (see \ref{sect:3c}) some
modifications have to be introduced in \eq{corr_fn}. Namely
\be
\chi(\mbf{r},\mbf{r}')^{-1}=\frac{\rho_0\,
\delta(\mbf{r}-\mbf{r}')}{\rho(\mbf{r})(\rho_0-\rho(\mbf{r}))}+\beta
V(\mbf{r}-\mbf{r}') \label{corr_fn2}
\ee
while the equation \eq{exp_corr_func} remains without any changes.

\section{From microscopic approach to Ginzburg--Landau--Chandler--type theory}
\label{sect:3}

In principle, the partition function \eq{PF_particle} describes all physical
properties of a liquid, including the possible phase transitions, while the
mean--field free energy functional \eq{free_energy} has a single minimum and hence
does not describe phase transition. This contradiction is due to the point--like
nature of particles in our treatment. We can overcome this obstacle introducing a
finite size for each particle. This is done in the Section \ref{sect:3c}. In the
meantime, in Sections \ref{sect:3a}, \ref{sect:3b} we split the free energy
functional \eq{free_energy} in long-- and short--ranged parts and derive the
solvation free energy using the technique of constraints in the functional integral.

\subsection{Expansion of the free energy}
\label{sect:3a}

Consider a solution $\bar{\rho}(\mbf{r})\equiv n(\mbf{r})$ describing an equilibrium
liquid--vapor interface. Supposing that the continuous profile $n(\mbf{r})$ is
rippled by fluctuations \cite{Weeks,Weeks_Bedeaux}, write the instantaneous value of the density as a sum of
two parts: $\rho(\mbf{r})= \omega(\mbf{r})+ n(\mbf{r})$, where $n$ corresponds to an
equilibrium interface and $\omega$ -- to fluctuations. Expand now the grand
thermodynamic potential near the profile $n(\mbf{r})$ in $\omega(\mbf{r})$ up to the
second order. This is possible if the fluctuations only slightly deform the
equilibrium profile. Thus, we have
\be
\Omega[\rho]=\Omega[n]+\int \left.\frac{\delta \Omega}{\delta
\rho(\mbf{r})}\right|_{\rho=n} \omega(\mbf{r})\, d\mbf{r} +\frac{1}{2}\int
\left.\frac{\delta^2 \Omega}{\delta \rho(\mbf{r})\delta
\rho(\mbf{r}')}\right|_{\rho=n} \omega(\mbf{r})\omega(\mbf{r}')\, d\mbf{r}d\mbf{r}'
\label{free_expansion}
\ee
By definition:
\be
\left.\frac{\delta \Omega}{\delta \rho(\mbf{r})}\right|_{\rho=n}=0, \qquad
\left.\frac{\delta^2 \Omega}{\delta \rho(\mbf{r})\delta \rho(\mbf{r}')}\right|_{\rho=n}
=\chi^{-1}(\mbf{r}-\mbf{r}') \label{deriv}
\ee
Substituting \eq{deriv} into \eq{free_expansion}, we get
\be
\Omega[\rho]=\Omega[n]+\frac{1}{2}\int \omega(\mbf{r}) \chi^{-1}(\mbf{r}-\mbf{r}')
\omega(\mbf{r}') d\mbf{r}d\mbf{r}' \label{FE_exp}
\ee
The density distribution $n(\mbf{r})$ is long--ranged, so its Fourier image
$n(\mbf{k})$ is nonzero at small $\mbf{k}$. Using this fact we can simplify
\eq{FE_exp} in the following manner. Consider the part of the Fourier space where
$n(\mbf{k})\neq 0$. In this region $V(\mbf{k})$ can be expanded in a Taylor series:
$V(\mbf{k})\simeq v_0+k_i\partial_iV(\mbf{k})+\frac{1}{2}k_i k_j\partial_i\partial_j
V(\mbf{k})$ ($\{i,j\}=1,2,3$). Under the symmetry arguments the last expression
reduces to
$$
V(\mbf{k})\simeq v_0+\frac{1}{2}v_2 \mbf{k}^2
$$
So, we get
$$
\int n(\mbf{r}) V(\mbf{r}-\mbf{r}') n(\mbf{r}') d\mbf{r}' d\mbf{r}=v_0 \int
n^2(\mbf{r}) d\mbf{r}+\frac{v_2}{2}\int \left(\nabla
n(\mbf{r})\right)^2 d\mbf{r}
$$
where
\be
v_0=\int V(\mbf{r}) d\mbf{r},\qquad v_2=-\int \mbf{r}^2 V(\mbf{r}) d\mbf{r}
\label{v_expansion}
\ee
Thus, $\Omega[n]$ appears as:
\be
\Omega[n]=\frac{1}{\beta}\int n(\mbf{r})(\ln n(\mbf{r})-1)
d\mbf{r}+\frac{v_0}{2}\int n^{2}(\mbf{r}) d\mbf{r}+\frac{v_2}{4}\int (\nabla
n(\mbf{r}))^2 d\mbf{r}
\ee
Collecting all terms we obtain a mean--field Ginzburg--Landau--Chandler--type
functional
\be\label{fr_en2}
\barr{lll} \disp \Omega=\frac{1}{2}\int \omega(\mbf{r}) \chi^{-1}(\mbf{r}-\mbf{r}')
\omega(\mbf{r}') d\mbf{r}d\mbf{r}' & + & \disp  \int \left\{\frac{v_2}{4}(\nabla
n(\mbf{r}))^2 + \tilde{W}(n({\bf r})) \right\}\, d\mbf{r}
\medskip \earr
\ee
where
\be
\tilde{W}(n({\bf r}))=\frac{v_0}{2} n^{2}(\mbf{r})+ \frac{1}{\beta} n(\mbf{r})(\ln
n(\mbf{r})-1) -\mu \,n(\mbf{r}) \label{poten}
\ee
One can clearly see the similarity between \eq{fr_en2}--\eq{poten} and
\eq{1}--\eq{eq:GLP}. As it has been mentioned in the Introduction, the precise form
of the potential $W(n({\bf r}))$ is not important -- it should only allow the phase
transition. Despite $\tilde{W}(n({\bf r}))$ does not fit this condition, its simple
generalization obtained in Section \ref{sect:3c} in the frameworks of the same
formalism satisfies all necessary requirements. The interaction term between
fluctuations of short-- and long--ranged fields identically vanishes in \eq{fr_en2},
so the coupling constant $c$ in \eq{1} should be assigned to zero.

The difference between the microscopic approach considered in this paper and the
approach based on scale separation \cite{Nechaev&Sitnikov&Taran} becomes much more
essential when we put the solute molecule into the solvent described by the
Hamiltonian \eq{mic_ham} and apply the technique of functional constraints as it has
been proposed by Li and Kardar \cite{Li&Kadar} and then exploited by Chandler {\em
et al} \cite{chand_gauss,chand_LCW,chand_lattice}.

\subsection{The cavitation part of the solvation free energy}
\label{sect:3b}

The influence of cavities, extended surfaces and other inhomogeneities of geometric
origin immersed in fluctuating environment, can be taken into account by means of a
technique \cite{chand_gauss} of constraints in functional integration. In brief, the
presence of a constraint in a partition function generates an artificial "ghost"
field coupled with other "real" fields describing the fluctuating media. This
interaction corresponds to the media (solvent) response on a constraint. In
particular, the cavitational part of a hydrophobic effect appears due to the
requirement to have no solvent particle inside the fixed cavity (solute molecule).

In this Section we start with the microscopic Hamiltonian \eq{mic_ham} and pass to
the collective variables (as it has been done in section \ref{sect:2}) demanding
that no particle enter the volume $V_{\rm in}$ of a solute molecule. The
corresponding partition function can be written as follows
\be
Z=\int \mcl{D}\{\rho\} \mcl{D}\{\phi\} \prod_{\mbf{r}\in v_{\rm in}} \delta[\rho]\,
e^{-S\{\rho,\phi\}} \label{pf0}
\ee
where the action $S$ has the form
\be
S\{\rho,\phi\}=\frac{\beta}{2} \int \rho(\mbf{r})V(\mbf{r}-\mbf{r}') \rho(\mbf{r}')
d\mbf{r}d\mbf{r}'-\beta \int \rho(\mbf{r}) \Big(\phi(\mbf{r})-U(\mbf{r})\Big)
d\mbf{r} -\lambda \int e^{-\beta \phi(\mbf{r})}d\mbf{r}
\ee
Using again the functional Fourier representation of $\delta$--function and rotation of $\psi$ to the imaginary axis, we get
$$
\barr{l} \disp Z=\mcl{N}_1  \int \mcl{D}\{\phi\} \mcl{D}\{\psi\}
\exp\left\{\frac{\beta}{2}\int d\mbf{r}d\mbf{r}'
\Big(\phi(\mbf{r})-U(\mbf{r})+\tau(\mbf{r}) \psi(\mbf{r})\Big)
V^{-1}(\mbf{r}-\mbf{r}')\Big(\phi(\mbf{r}')-U(\mbf{r}')+ \tau(\mbf{r}')
\psi(\mbf{r}')\Big)\right. \\ \disp \hspace{4cm} \left.+\lambda \int d\mbf{r}
e^{-\beta \phi(\mbf{r})}\right\} \earr
$$
Integrating over the fields $\rho$ and $\psi$, we arrive at the following expression
\begin{align}
Z=\mcl{N}_2 \int & \mcl{D}\{\phi\} \exp\left\{\frac{\beta}{2}\int d\mbf{r} d\mbf{r}'
\Big(\phi(\mbf{r})-U(\mbf{r})\Big)V^{-1}(\mbf{r}-\mbf{r}')
\Big(\phi(\mbf{r}')-U(\mbf{r}')\Big)+\lambda \int d\mbf{r}
e^{-\beta \phi(\mbf{r})}\right\} \nonumber \\ & \times
\exp\left\{-\frac{\beta}{2} \int d\mbf{r} d\mbf{r}' d\mbf{r}''d\mbf{r}'''
\Big(\phi(\mbf{r})-U(\mbf{r})\Big) V^{-1}(\mbf{r}-\mbf{r}'')V_{\rm in}
(\mbf{r}''-\mbf{r}''') V^{-1}(\mbf{r}'''-\mbf{r}')\Big(\phi(\mbf{r})'-
U(\mbf{r}')\Big)\right\} \label{pf}
\end{align}
here
\be
\barr{rcll} \disp \int_{v_{\rm in}}\, d{\bf r}''V_{\rm in}({\bf
r},{\bf r}'') V({\bf r}''-{\bf r}') & = & \delta({\bf r}'-{\bf r}) & \qquad
\mbox{for ${\bf r},{\bf r'}\in v_{\rm in}$} \medskip \\ \disp V_{\rm in}({\bf r},{\bf
r}'') & = &  0 & \qquad \mbox{for ${\bf r}\in v_{\rm out}$} \earr .
\label{chi_inv}
\ee
and integration in \eq{pf} on $\mbf{r}''$ and $\mbf{r}'''$ actually is carried out in volume $v_{\rm in}$.

Introducing the new variable
\be
\vartheta(\mbf{r})=\int V(\mbf{r}-\mbf{r}')\Big(\phi(\mbf{r}')-U(\mbf{r}')\Big)
d\mbf{r}'
\ee
the partition function \eq{pf} can be rewritten as
\be
Z=\mcl{N}_3\int \mcl{D}\vartheta\, e^{-\tilde{S}\{\vartheta\}}
\ee
where
\be
\barr{lll} \disp \tilde{S}\{\vartheta\}=\frac{\beta}{2} \int \vartheta(\mbf{r})
V(\mbf{r}-\mbf{r}') \vartheta(\mbf{r}') d\mbf{r} d\mbf{r}' & - & \disp
\frac{\beta}{2} \int \vartheta(\mbf{r}) V_{\rm in}(\mbf{r},\mbf{r}')
\vartheta(\mbf{r}') d\mbf{r} d\mbf{r}' \\ & + & \disp \lambda \int
\exp\left\{-\beta\left(U(\mbf{r})+\int V(\mbf{r}-\mbf{r}')\,\vartheta(\mbf{r}')
d\mbf{r}'\right)\right\} \earr \label{varphi_avr}
\ee
In a saddle point approximation we have
\be
\barr{lll} \disp \frac{\delta \tilde{S}\{\vartheta\}}{\delta \vartheta}=\beta\int
V(\mbf{r}-\mbf{r}') \bar{\vartheta}(\mbf{r}') d\mbf{r}' & - & \disp \beta \int
V_{\rm in}(\mbf{r},\mbf{r}') \bar{\vartheta}(\mbf{r}') d\mbf{r}' \\ & - & \disp
\beta \lambda\int V(\mbf{r}-\mbf{r}') \exp\left\{-\beta\left(U(\mbf{r}')+\int
V(\mbf{r}'-\mbf{r}'')\bar{\vartheta}(\mbf{r}'') d\mbf{r}''\right)\right\}
d\mbf{r}'=0 \earr \label{varphi_avr2}
\ee
Calculating now the mean density $\bar{\rho}(\mbf{r})$ in a standard manner, we
arrive at the following expression
\be
\bar{\rho}(\mbf{r})=\lambda \exp\left\{-\beta \left(U(\mbf{r})+\int
V(\mbf{r}-\mbf{r}')\bar{\vartheta}(\mbf{r}') d\mbf{r}'\right)\right\}
\label{rho_mean2}
\ee
where $\bar{\vartheta}(\mbf{r})$ is defined from \eq{varphi_avr2}
\be
\bar{\vartheta}(\mbf{r})-\int V^{-1}(\mbf{r}-\mbf{r}') V_{\rm
in}(\mbf{r}',\mbf{r}'')\bar{\vartheta}(\mbf{r}'') d\mbf{r}''=\lambda
\exp\left\{-\beta\left(U(\mbf{r})+\int V(\mbf{r}-\mbf{r}') \bar{\vartheta}(\mbf{r}')
d\mbf{r}'\right)\right\} \label{varphi_avr3}
\ee

Let us pay attention to the equations \eq{rho_mean2}--\eq{varphi_avr3} obtained
using the standard technique of functional constraints with the microscopic
Hamiltonian {\it before} separation of the density field $\rho({\bf r})$ into
short-- and long--ranged scales. The equation \eq{varphi_avr3} is contradictory to
some extent. Namely, for all $\mbf{r}\in v_{\rm in}$ the left--hand side of
\eq{varphi_avr3} identically vanishes, while the right--hand side is positive. This
contradiction can be formally removed by noting that the potential $U$ creates an
impenetrable wall around the volume $v_{\rm in}$. However, in this case the value of
the field $\bar{\vartheta}$ is undetermined inside the volume $v_{\rm in}$.

So, the account of hydrophobic effect by natural inserting constraint into the
functional integral \eq{pf0} leads to internal inconsistency in the two--length
scale theory. One can overcome this contradiction by using the two--length scale
expansion \eq{fr_en2}--\eq{poten} {\it on the first step}, and applying the
functional constraint {\it on the subsequent step}, against the direct use of
constraint on microscopic level as in \eq{pf0}, however such a procedure looks a bit
artificial.

\subsection{Solvent particles of finite volume}
\label{sect:3c}

The simplest way to take into account the finiteness of the volume occupied by
particles of the solvent consists in introducing an additional artificial density
$\rho_{\rm v}({\bf r})$ associated with voids between particles, and requiring then
the sum of two densities, $\bar{\rho}({\bf r})$ and $\rho_{\rm v}({\bf r})$ to be
constant everywhere \cite{Kholodenko&Qian}. Such a condition guarantees that when a
particle leaves some volume, it is replaced by a "hole". Such consideration is a
sort of an off--lattice Flory--Huggins mean--field treatment. Following the same
steps as in the Section \ref{sect:2}, we arrive finally at the free energy
functional, which generates the grand potential functional with a double--well
structure, describing a phase transition:
\be
F=\frac{1}{2}\int \bar{\rho}(\mbf{r}) V(\mbf{r}-\mbf{r}')
\bar{\rho}(\mbf{r}')d\mbf{r}d\mbf{r}'+\frac{1}{\beta}\int
\bar{\rho}(\mbf{r})\ln\bar{\rho}(\mbf{r}) d\mbf{r} + \frac{1}{\beta}\int
\Big(\rho_0-\bar{\rho}(\mbf{r})\Big)\ln\Big(\rho_0-\bar{\rho}(\mbf{r})\Big)
d\mbf{r}+\int U(\mbf{r}) \bar{\rho}(\mbf{r}) d\mbf{r}\label{free_energy2}
\ee
where $\rho_{\rm v}(\mbf{r})=\rho_0-\bar{\rho}(\mbf{r})$ and $\rho_0$ is the inverse
volume of a media molecule. Rewriting the entropy term, neglecting the meaningless
constants and proceeding to the grand potential $\Omega$, we get:
\be
\Omega=\frac{\rho_0}{\beta}\int
\frac{\bar{\rho}(\mbf{r})}{\rho_0}\ln\frac{\bar{\rho}(\mbf{r})}{\rho_0}
+\left(1-\frac{\bar{\rho}(\mbf{r})}{\rho_0}\right)
\ln\left(1-\frac{\bar{\rho}(\mbf{r})}{\rho_0}\right) d\mbf{r}+\frac{1}{2}\int
\bar{\rho}(\mbf{r}) V(\mbf{r}-\mbf{r}') \bar{\rho}(\mbf{r}')d\mbf{r}d\mbf{r}'-\int
\mu \bar{\rho}(\mbf{r}) d\mbf{r}+\int U(\mbf{r}) \bar{\rho}(\mbf{r})
d\mbf{r}\label{free_energy3}
\ee

The equilibrium density profile is determined by the condition
\be
\frac{\delta \Omega}{\delta \rho(\mbf{r})}=0
\ee
which leads to the following equation:
\be
\bar{\rho}(\mbf{r})=\Big(\rho_0-\bar{\rho}(\mbf{r})\Big)\exp\left\{\beta
\left(\mu-U(\mbf{r})-\int V(\mbf{r}-\mbf{r}')\bar{\rho}(\mbf{r}')
d\mbf{r}'\right)\right\} \label{rho_mean3}
\ee
Let us focus attention on the structure of $\Omega$. There are two independent
parameters, namely $\beta$ and $\mu$, which determine the thermodynamic state of the
system. In a homogeneous media the density of the grand thermodynamic potential may
be written as
\be
-p=\frac{\bar{\rho}}{\beta} \ln\frac{\bar{\rho}}{\rho_0} +\rho_0
\left(1-\frac{\bar{\rho}}{\rho_0}\right)\ln\left(1-\frac{\bar{\rho}}{\rho_0}\right)
+\frac{1}{2}v_0\, \bar{\rho}^2- \mu \bar{\rho}
\label{eq:equil}
\ee
The quantity $v_0$ is determined by \eq{v_expansion}. It is easy to see that
\eq{eq:equil} describes the phase coexistence when $\mu_c =\rho_0 v_0/2$. Then the
critical temperature $T_c$ of the vapor--liquid phase transition is determined by
$\partial^2 p/\partial \rho^2=0$, giving $T_c=-v_0 \rho_0/4$. Below $T_c$ there are
two different phases, while above $T_c$ the system is homogenious. Other values of
$\mu$ correspond to cases of single phase in the system, vapor or liquid. The value
of $\rho_0$ is fixed by the bulk density $\rho_{\rm b}$. Sufficiently far from the
solvated object $U(\mbf{r})\simeq 0$, thus we have:
\be
\ln\left(\frac{\rho_{\rm b}}{\rho_0-\rho_{\rm b}}\right)=
\beta v_0\left(\frac{\rho_0}{2}-\rho_{\rm b}\right)
\label{constants}
\ee
As was mentioned above the parameter $\rho_0$ is related to the inverse volume of
the molecule. The latter is rather arbitrary, because it strongly depends on the
convention how the intra--molecular potential is divided into repulsive and
attractive parts. Let us stress that in our approach we do not use explicitly the
pairwise potential of liquid molecules $V({\bf r}-{\bf r}')$, it enters effectively
only through the correlation function $h({\bf r}-{\bf r}')$---see
\eq{corr_fn}--\eq{exp_corr_func}. For example, we can associate the parameter
$\rho_0$ with the repulsive part of potential and the $V(\mbf{r}-\mbf{r}')$ -- with
the smooth attractive part. So, one can say that the relation \eq{constants} fixes
the value of $\rho_0$, while the correlation function of the liquid $h({\bf r}-{\bf
r}')$ fixes the potential $V(\mbf{r}-\mbf{r}')$ via \eq{exp_corr_func}.

\section{Discussion}
\label{sect:4}

The aim of the present paper is two--fold. On one hand, we have derived the
self--consistent set of equations which determine the solvation free energy of a
molecule of arbitrary shape immersed into a solution. The shape of a solute molecule
is completely defined by the collection of its Van-der-Waals radii and is encoded in
the potential $U({\bf r})$ coupled to the equilibrium solvent density
$\bar{\rho}({\bf r})$. In turn, the equilibrium density $\bar{\rho}({\bf r})$ is
determined by the nonlinear equation containing the pairwise density--density
correlation function, $\chi({\bf r}-{\bf r}')$ of a pure solvent, which is supposed
to be known and serves as an input in the theory. The full set of corresponding
equations \eq{free_energy3}, \eq{rho_mean3}, \eq{corr_fn} is collected below for
convenience:
\be
\left\{\barr{l} \disp F=\frac{1}{2}\int \bar{\rho}(\mbf{r}) V(\mbf{r}-\mbf{r}')
\bar{\rho}(\mbf{r}')d\mbf{r}d\mbf{r}'+\frac{1}{\beta}\int
\bar{\rho}(\mbf{r})\ln\bar{\rho}(\mbf{r}) d\mbf{r} + \frac{1}{\beta}\int
\Big(\rho_0-\bar{\rho}(\mbf{r})\Big)\ln\Big(\rho_0-\bar{\rho}(\mbf{r})\Big)
d\mbf{r}+\int U(\mbf{r}) \bar{\rho}(\mbf{r}) d\mbf{r} \medskip \\
\disp \bar{\rho}(\mbf{r})=\Big(\rho_0-\bar{\rho}(\mbf{r})\Big)
\exp\left\{\beta \left(\frac{\rho_0}{2}\left[\int V({\bf r})\, d{\bf r}\right]
-U(\mbf{r})-\int V(\mbf{r}-\mbf{r}')\bar{\rho}(\mbf{r}')
d\mbf{r}'\right)\right\} \medskip \\
\disp V(\mbf{r}-\mbf{r}')= \frac{1}{\beta}\, \chi(\mbf{r},\mbf{r}')^{-1} -
\frac{\rho_0\, \delta(\mbf{r}-\mbf{r}')}{\beta\,
\rho(\mbf{r})(\rho_0-\rho(\mbf{r}))}
\earr \right. \label{fin}
\ee
The correlation function $\chi({\bf r}-{\bf r}')$ and the density $\rho_{\rm b}$ in
the bulk  enter as inputs in the theory, while the relation between the constants
$\rho_0$ and $\rho_{\rm b}$ is established in \eq{constants}. Let us stress that the
free energy of interactions between different solute molecules can be
straightforwardly computed on the basis of \eq{fin}.

On the other hand, we have established the internal inconsistency between the
microscopic approach and the two--ranged scale Ginzburg--Landau--Chandler (GLC)
description of hydrophobic effect.

Let us remind that the Hamiltonian of the GLC--type theory consists of two
phenomenological parts: (i) the nonlocal Gaussian term corresponding to
short--ranged interactions between pure solvent molecules in the bulk, and (ii) the
"long--ranged" term coming from smoothly changing profile of "macroscopic" (i.e.
averaged over short--ranged fluctuations) solvent density taken in a form of a
Ginzburg--Landau expansion. These two contributions to the free energy are decoupled
until the solute molecule is put into a solvent. In the frameworks of GLC theory the
solute expels the total solvent density $\rho({\bf r})$ from the volume. As soon as
$\rho({\bf r})$ is a sum of short-- and long--ranged contributions, the condition
$\rho({\bf r})=0$ inside the solvent molecule leads to an effective coupling between
short-- and long--ranged fields. Then one can proceed with the standard
thermodynamic formalism and compute the averaged density, the solvation free energy,
etc.

The origin of the discrepancy between microscopic approach developed above and
GLC--type theory consists in the following. If we use the functional constraint
nullifying the total solvent density $\rho$ (as in GLC theory) in the partition
function with microscopic Hamiltonian (see eq.\eq{pf0}), then we come to the
contradiction: some fields entering in the answer cannot be accurately defined (for
more details see the last paragraph of Section \ref{sect:3b}). Only for pure solvent
the decomposition on short-- and long--ranged scales (as in GLC approach) is
consistent with microscopic description---compare \eq{fr_en2}--\eq{poten} and
\eq{1}--\eq{eq:GLP}. At the same time the approach described in the current paper
does not require any coupling between fields describing the microscopic and
macroscopic structures of the solvent---in our work we get rid of any artificial
division of fields in different scale ranges. Moreover, the values of the
coefficients $a$ and $b$ in the GLC Hamiltonian are not arbitrary, but are dictated
by equation \eq{v_expansion}.

Meanwhile, we consider it necessary to note that still the two--length scale model,
being properly tuned, provides rather successful quantitative description of
cavitation part of hard spheres and the solvation energies of alkane molecules as it
has been shown recently, for example, in \cite{Nechaev&Sitnikov&Taran}. So, to our
point of view the GLC--type approach is nevertheless satisfactory as a
phenomenological theory.

\begin{acknowledgments}
We are very grateful to M. Taran for valuable comments. One of us (G.S.) would thank
the laboratory LPTMS (Universit\'e Paris Sud, Orsay) for the financial support
during the period in which the work has been started.

\end{acknowledgments}

\end{document}